# JUNO DAQ Readout and Event Building Research

Tingxuan Zeng, Fei Li, Kejun Zhu

*Abstract*—The Jiangmen Underground Neutrino Observatory(JUNO) experiment will develop an internationally leading neutrino experimental station to measure the order of neutrino mass and carry out many other scientific frontier studies. The entire experimental system includes the detector system, the electronics system, and the data acquisition (DAQ) system. Data flow is the core of JUNO DAQ system. Readout system and Event Building (EB) system are two key aspects of data flow. Based on the requirements analysis of JUNO DAQ, this paper proposes a data flow schema of distributed network readout and second-level event building. Focusing on the performance requirements of JUNO DAQ, the performance of the readout and event building module, the number of readout processes deployed on each node, and integration performance of the two modules are studied. The results of the research provide a reference for further optimization of the JUNO DAQ data flow framework.

Keywords：JUNO DAQ, data flow, event building, readout, performance

## I. INTRODUCTION

THE Jiangmen Underground Neutrino Observatory(JUNO) experiment will develop an internationally leading neutrino experimental station to measure the order of neutrino mass and carry out many other scientific frontier studies. The entire experimental system includes the detector system, the electronic system, and the data acquisition (DAQ) system. The detector system detects signal , the electronics system converts the signal into binary data, and the DAQ system receives, processes and stores the binary data. The detector system consists of three detectors with a total of nearly fifty thousand PMTs, of which about twenty thousand PMTs are required for full-channel 1GHz FADC waveform sampling [1]. The sampling window is planned to be 1us. Each sampling point occupies about 16bits, so single PMT signal data length will reach 2KB. After triggering, the event rate will decreases to 1KHz. Without any compression on the data channels, the corresponding readout throughput will be 40GB/s. If a one-to-one detector and electronics connection scheme is used, there will be about twenty thousand read-out connections [2]. In order to read out and process about twenty thousand channels FEE data, DAQ readout module will use distributed network readout schema. The DAQ system needs to assemble the data fragments of all electronic channels into a full event of the detector according to the trigger id.

## II. READOUT AND EVENT BUILDING SCHEMA IMPLEMENTATION

Since FEE don't have flexible data scheduling transmission capability, The event building process will be scheduled after the DAQ readout module. The readout module assembles multiple channels of electronics data together to form event fragments, and then sends the fragments to event building module. The event building module assembles these fragments to form a full event.

JUNO experiment has many similarities with BESIII、ATLAS and Daya Bay experiments in DAQ data flow section [3][4]. We can design and develop JUNO DAQ data flow software by referring to the three experiments mentioned above. In JUNO DAQ data flow framework, the readout process of the readout module is called ROS, and the event building process in event building module is called EB. The readout module reads out the electronic data and performs the first-level data assembly, the event building module collects all ROS fragments and packages them into full event. ROS module defines abstract readout interface. Different implementation of the readout interface corresponds to different ways to read out FEE data. According to the front-end electronic interface, we implement a client interface for network readout and integrate it into the ROS module .

Fig. 1 describes the data flow collaboration process for readout and event building module [4]. It mainly includes the read-out subsystem(ROS), event building system(EB), Event Building Manager(EBM). A ROS receives multiple channels of fragments from the FEE. After collecting all the fragments of a specific event id, ROS sends the event id(L1id) to EBM. The EBM assigns the event to an unbusy EB via a "round robin" load balancing algorithm [5]. The assigned EB requests all the fragments of the event(event L1id) from all the ROSs and builds a full event out of them . After the EB has completed building an event, it sends an 'end of event' message back to the EBM. The EBM then removes the event from its internal bookkeeping and sends the clear command to the ROSs.

Tingxuan Zeng is with the State Key Laboratory of Particle Detection and Electronics, IHEP, CAS, UCAS, Beijing 100049, China (e-mail: zengtx@ ihep.ac.cn).
Fei Li is with the State Key Laboratory of Particle Detection and Electronics, IHEP, CAS, UCAS, Beijing 100049, China  (e-mail: lifei@ ihep.ac.cn).
Kejun Zhu is with the State Key Laboratory of Particle Detection and Electronics, IHEP, CAS, UCAS, Beijing 100049, China (e-mail: zhukj@ ihep.ac.cn).

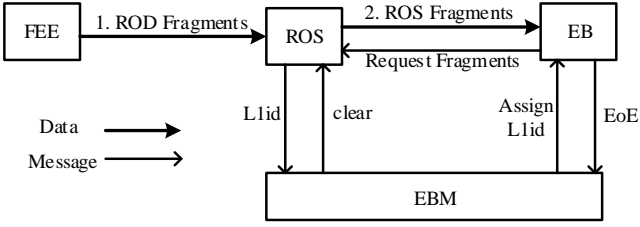

Fig. 1. JUNO DAQ readout and event building data flow collaboration diagram.

After receiving the clear messages, all the ROSs then delete the corresponding fragments from its memory. The performance of the read-out module corresponds to the bandwidth performance of process 1 in figure 1.1, and the EB performance corresponds to the bandwidth performance of process 2.

## III. RESEARCH TEST PLATFORM

In order to study the performance of the readout module, the front-end electronics (FEE) data source uses a server-side program developed by C ++ to send simulated data as fast as possible. All software is deployed on a set of Lenovo blade servers consisting of 14 nodes, each node containing a 56Gb / s IB NIC and a 10Gb Ethernet card. Other detailed parameters are shown in TABLE I :

TABLE I
RESEARCH TEST ENVIRONMENT MAIN PARAMETERS

|  | Blade node |
|---|---|
| CPU model | Intel(R) Xeon(R) CPU E5-2620 v2 @ 2.10GHz |
| CPU cores | 12 |
| Hyperthreading | on |
| OS | Scientific Linux CERN SLC release 6.6 |
| Kernel version | 2.6.32-504.el6.x86_64 |

## IV. READOUT MODULE PERFORMANCE STUDY

### A. Single readout node deployment

Single readout node bandwidth (Bandwidth) equals to the product of the fragment size(Size), the event rate(Rate) , the number of channels received per ROS(Chan) and the number of ROSs arranged on a node (NROS), As shown in the following formula:

$$Bandwidth = Size * Rate * Chan * NROS \quad \cdots\cdots (1)$$

According to the first chapter of the requirement analysis, JUNO DAQ's current performance design indexes can be summarized as:

$$Size = 2KB,\ Rate = 1KHz \quad \cdots\cdots (2)$$

The bandwidth and CPU resources of a single readout node are limited. The meaning of optimal deployment can be defined as : Under the condition of (2), adjust the value of NROS and Chan to reach maximum Bandwidth.

Fig. 2 is a set of tests conducted on a set of Lenovo blades using third-party software iperf, where the abscissa is the logarithm of threads number and the ordinate is the bandwidth. The fragment size is set to 2KB. Different lines represent different number of sending and receiving nodes, purple represents 1 sending node vesus 1 receiving node, green represents 2 sending nodes vesus 1 receiving node and blue represents 1 sending node vesus 2 receiving nodes. In the case of 1 vs 1, the purple line shows that 8 threads reach the maxmium bandwidth 27Gb/s at 2KB fragment length. Increasing a sending node, the green line shows almost the same bandwiths at different thread number, which indicates the bottleneck is not in sending node as performance doesn't increase while adding more sending cpu resource. In the case of 1 sending node vesus 2 receiving nodes, it reaches the maximum bandwidth 44Gb/s at 16 threads.This shows that the sending capability of a single node is greater than the receiving capability, and the cpu of the receiving node is a bottleneck.

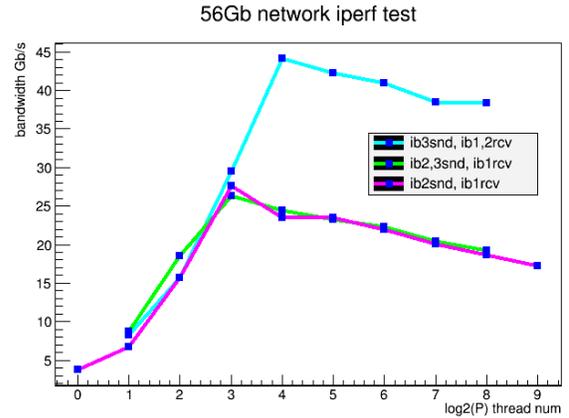

Fig. 2. iperf 56Gb network performance ,Size = 2KB

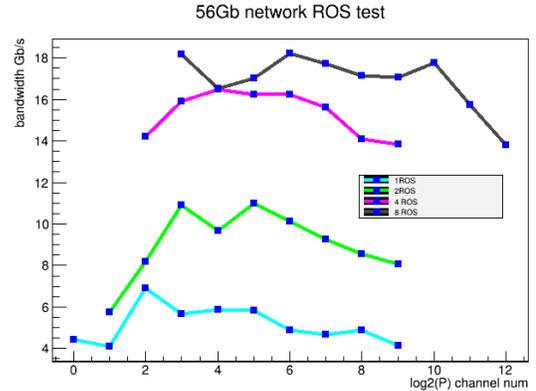

Fig. 3. ROS 56Gb network single readout node performance

Fig. 3 shows the bandwidth of single readout node as a function of the number of channels received per ROS（Chan）and the number of ROSs arranged on the node (NROS). In this set of tests ROS disabled memory management, message processing thread RequestHandler and other code has nothing to do with the readout process. It can be seen from figure 4.2 that best performance 18Gbps can be obtained when 8 ROSs are arranged on single node, supporting for 1024 channels. Performance doesn't increase when arranged more than 1024 channels.

### B. Scalability of multiple readout nodes

The scalability of software can be verified by extending the deployment scheme of single node to multiple nodes. Fig. 4 shows the bandwidth performance on each node when the single readout node deployment is extended to six nodes. With single receive buffer (which was used in previous sections), 21Gb/s bandwidth can be achieved per node. The socket buffer is set to 8MB to improve the readout performance. It shows that readout module has good scalability.

In order to insert, search and delete data fragments quickly, the daq software had its own memory management —— ROS Buffer Management [5]. While using ROS Buffer Management, with the same deployment on each node, less performance would be reached.

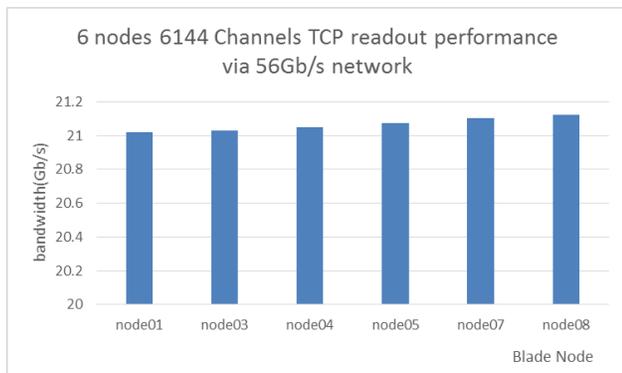

Fig. 4. Multiple nodes TCP readout performance with single receive buffer

## V. EVENT BUILDING MODULE PERFORMANCE

To reduce the burden on readout CPU and network, the ROS is replaced by EmuROS, which generates local dummy ROS fragments and sends them to the event building module. The number of EmuROSs arranged per node is called NEmuROS. The number of channels handled per EmuROS is called EmuChan.

Apply the optimal single-node deployment in the previous chapter, NROS = 8 and Chan = 128, to the simulated readout node, i.e. NEmuROS = 8, EmuChan = 128. The simulated readout node generates locally 2 KB of dummy fragments and sends them to the EB node. As the number of EBs increases from 1 to 4, the event rate remains constant at 0.7 kHz, which is less than the designed value of 1KHz. Network bandwidth is not full, adding CPU resource of the EB node does not help with performance improvement, indicating that the bottleneck at this time is in the CPU of the simulated readout module. In order to meet the data rate requirements, the number of data channels handled per node should be reduced. Fig. 5 shows 8 simulated readout nodes vesus 6 event building nodes performance with 384 channels deployed on each readout node.

The achieved event rate is about 1.25KHz, over the required 1KHz event rate.

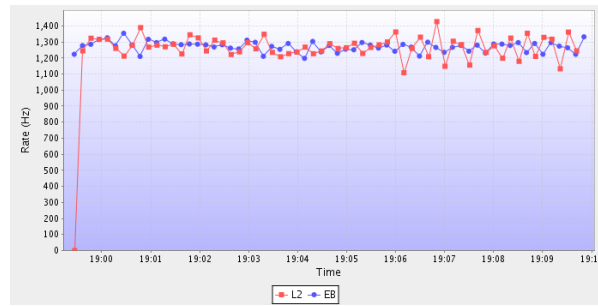

Fig. 5. 8 simulated readout node vs 6 EB node performance

## VI. READOUT AND EB INTEGRATION TEST RESULTS

Integrating the EB module and the readout module together with the simulated FEE data source, we get a set of software for integration testing. On the existing 14 blade servers: 2 sending nodes, 768 sending processes on each sending node; 4 readout nodes, 4 EB nodes are deployed. 1536 channels have been handled. The integration test results are shown in Fig. 6. It shows that the overall integration event rate is 1.1KHz. With 4 readout nodes, 4 EB nodes , 1536 channels of electronic data in total can be read out and assembled.

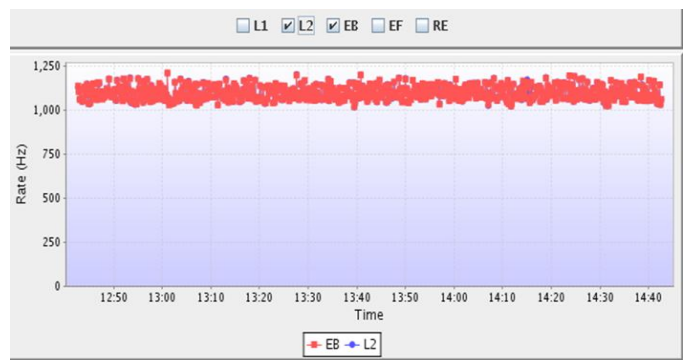

Fig. 6. Readout and EB dataflow integration test results

## VII. SUMMARY

The readout performance of JUNO DAQ is mainly affected by the number of readout processes, the number of channels connected to each readout process, and the fragment size. When fragment size is set to 2 KB, with 8 readout processes per node, 128 channels per process, we can reach a readout bandwidth of 17.7Gb/s, supporting up to 1024 channels per node. This is the optimal readout deployment with single receiving buffer and it has good scalability. In terms of the ROS buffer management, the readout performance decreases significantly, which can support up to 384 channels per readout node.

For the EB module, The assembly capability of a single EB node is stronger than the sending capability of a single simulated readout module. The deployment: Size=2KB,NROS=16, NChan=24, #EBs/Node=6 , can achieve en event rate of 1.25KHz, which preserves some event rate margins.

The integrated testing results show that 8 nodes can readout and assemble 1536 channels of electronic data with an event rate of 1.1KHz. Expanding to 20 thousand channels, a total of 105 such nodes, i.e. eight sets of blade servers are needed to complete the readout and event building process.